\documentclass[aps,prb,twocolumn,superscriptaddress]{revtex4-2}

\usepackage{graphicx, color, xcolor}
\usepackage{amsmath,amsfonts,amssymb}
\usepackage{booktabs} %
\usepackage{soul}
\usepackage{hyperref}
\hypersetup{
    colorlinks=true,
    linkcolor=blue,
    filecolor=magenta,      
    urlcolor=cyan,
    pdftitle={A Cryogenic Muon Tagging System Based on Kinetic Inductance Detectors for Superconducting Quantum Processors},
    }
    
\usepackage[heightadjust=all]{floatrow}
\usepackage{comment}

\providecommand{\ignore}[1]{}

\def\lsim{\mathrel{\rlap{\lower4pt\hbox{\hskip1pt$\sim$}}
    \raise1pt\hbox{$<$}}}                
\def\gsim{\mathrel{\rlap{\lower4pt\hbox{\hskip1pt$\sim$}}
    \raise1pt\hbox{$>$}}}                

\begin{document}

\title{A Cryogenic Muon Tagging System Based on Kinetic Inductance Detectors for Superconducting Quantum Processors}

\author{Ambra Mariani}
\email{Corresponding author: ambra.mariani@roma1.infn.it}
\affiliation{INFN -- Sezione di Roma, Roma I-00185, Italy}

\author{Laura Cardani}
\affiliation{INFN -- Sezione di Roma, Roma I-00185, Italy}

\author{Mustafa Bal}
\affiliation{Superconducting Quantum Materials and Systems Center (SQMS), Fermi National Accelerator Laboratory, Batavia, IL 60510, USA}

\author{Nicola Casali}
\affiliation{INFN -- Sezione di Roma, Roma I-00185, Italy}

\author{Ivan Colantoni}
\affiliation{Consiglio Nazionale delle Ricerche, Istituto di Nanotecnologia}
\affiliation{INFN -- Sezione di Roma, Roma I-00185, Italy}

\author{Angelo Cruciani}
\affiliation{INFN -- Sezione di Roma, Roma I-00185, Italy}

\author{Giorgio Del Castello}
\affiliation{INFN -- Sezione di Roma, Roma I-00185, Italy}

\author{Daniele Delicato}
\affiliation{INFN -- Sezione di Roma, Roma I-00185, Italy}
\affiliation{Dipartimento di Fisica, Sapienza Universit\`a di Roma, Roma I-00185, Italy}
\affiliation{Univ. Grenoble Alpes, CNRS, Grenoble INP, Institut Néel, 38000 Grenoble, France}

\author{Francesco De Dominicis}
\affiliation{Gran Sasso Science Institute, L’Aquila I-67100, Italy}
\affiliation{INFN – Laboratori Nazionali del Gran Sasso, Assergi (L’Aquila) I-67100, Italy}

\author{Matteo del Gallo Raccagiovine}
\affiliation{INFN -- Sezione di Roma, Roma I-00185, Italy}
\affiliation{Dipartimento di Fisica, Sapienza Universit\`a di Roma, Roma I-00185, Italy}

\author{Matteo Folcarelli}
\affiliation{INFN -- Sezione di Roma, Roma I-00185, Italy}
\affiliation{Dipartimento di Fisica, Sapienza Universit\`a di Roma, Roma I-00185, Italy}

\author{Sabrina Garattoni}
\affiliation{Superconducting Quantum Materials and Systems Center (SQMS), Fermi National Accelerator Laboratory, Batavia, IL 60510, USA}

\author{Anna Grassellino}
\affiliation{Superconducting Quantum Materials and Systems Center (SQMS), Fermi National Accelerator Laboratory, Batavia, IL 60510, USA}

\author{Mehmood Khan Yasir Raja}
\affiliation{Gran Sasso Science Institute, L’Aquila I-67100, Italy}
\affiliation{INFN – Laboratori Nazionali del Gran Sasso, Assergi (L’Aquila) I-67100, Italy}

\author{Valerio Pettinacci}
\affiliation{INFN -- Sezione di Roma, Roma I-00185, Italy}

\author{Alberto Ressa}
\affiliation{INFN -- Sezione di Roma, Roma I-00185, Italy}

\author{Tanay Roy}
\affiliation{Superconducting Quantum Materials and Systems Center (SQMS), Fermi National Accelerator Laboratory, Batavia, IL 60510, USA}

\author{Marco Vignati}
\affiliation{INFN -- Sezione di Roma, Roma I-00185, Italy}
\affiliation{Dipartimento di Fisica, Sapienza Universit\`a di Roma, Roma I-00185, Italy}

\author{David van Zanten}
\affiliation{Superconducting Quantum Materials and Systems Center (SQMS), Fermi National Accelerator Laboratory, Batavia, IL 60510, USA}

\date{\today}

\begin{abstract}
\textbf{Abstract:} Ionizing radiation has emerged as a potential limiting factor for superconducting quantum processors, inducing quasiparticle bursts and correlated errors that challenge fault-tolerant operation. 
Atmospheric muons are particularly problematic due to their high energy and penetration power, making passive shielding ineffective. Therefore, monitoring the real-time muon flux is crucial to guide the development of alternative error-correction or mitigation strategies.

We present the design, simulation, and first operation of a cryogenic muon-tagging system based on Kinetic Inductance Detectors (KIDs), developed as a stand-alone cryogenic particle-tagging module for superconducting quantum processors. The system consists of two KIDs arranged in a vertical stack and operated at $\sim$20 mK. Monte Carlo simulations based on Geant4 guided the prototype design and provided reference expectations for muon-tagging efficiency and accidental coincidences due to ambient $\gamma$-rays.

We observed a muon-induced coincidence rate among the top and bottom detectors of (192 $\pm$ 9)$\times10^{-3}$\,events/s, in excellent agreement with the Monte Carlo prediction. The prototype achieves a muon-tagging efficiency of about 90\% with negligible dead time. 

These results demonstrate the feasibility of operating a muon-tagging system at millikelvin temperatures and represent a key step toward the integration of cryogenic veto systems with multi-qubit chips to mitigate muon-induced errors.
\end{abstract}

\maketitle
\section{Introduction}
\label{sec:introduction}
Superconducting quantum processors are among the most promising platforms for scalable quantum computation, offering design flexibility, high-fidelity operations, and compatibility with fast readout architectures~\cite{Barends2014, Kjaergaard2020}. 
Significant advancements in the past decade include the demonstration of quantum advantage~\cite{Arute:2019, Wu:2021, GoogleQuantumAI:2024, Gao:2025}, rapid progress in scaling up the number of qubits~\cite{Rigetti, IBM:eagle, IBM_processors, Google_Willow_chip}, and substantial improvements in qubit coherence times~\cite{Place:2021, Somoroff:2023, Grassellino2023, Bland:2025}. 
These developments are key milestones toward fault-tolerant quantum computing.

However, ionizing radiation has recently been recognized as a source of decoherence in superconducting qubits, limiting device performance~\cite{Vepsalainen:2020}. Energy deposited by environmental $\gamma$-rays or cosmic-ray muons in the chip substrate generates free charges, whose recombination produces high-energy phonons. These phonons propagate across the device, breaking Cooper pairs and producing bursts of quasiparticles. Quasiparticle tunneling across the Josephson Junction can significantly degrade the relaxation time of single qubits and, in some cases, induce correlated errors across multiple qubits on the same chip~\cite{Wilen:2021, McEwen:2021, Harrington:2024, Li2024direct}. These errors violate the assumptions of spatial and temporal independence that underlie most quantum error correction algorithms and represent a major challenge to large-scale quantum computing. Multiple studies have demonstrated a direct correlation between radiation exposure and qubit performance degradation, with event rates consistent with ambient radiation levels~\cite{Vepsalainen:2020, DeDominicis:2025}. Other investigations, however, suggest that ionizing radiation may also couple to two-level systems (TLSs)~\cite{Thorbeck:2022}, a known dominant noise source in superconducting circuits.

To mitigate these effects, a variety of strategies have been explored. Careful material selection, passive shielding, and the use of underground laboratories can reduce radiation-induced instabilities and improve superconducting device performance~\cite{Bratrud2024FNAL, Gusenkova:2022, Cardani:2021}. However, deploying superconducting processors in deep-underground facilities introduces substantial infrastructure complexity. 
Other promising approaches include phonon ``traps"~\cite{Henriques:2019, martinis:2021, iaia_2022}, gap engineering~\cite{mcewen2024gap_engineering, Kamenov:2024} to reduce quasiparticles tunneling, and even the integration of external radiation sensors~\cite{Orrell:2021}. While the interaction rate in a superconducting quantum chip is dominated by $\gamma$-rays from environmental radioactivity, which outnumber atmospheric muons by at least a factor of two~\cite{cardani:2023, Loer:2024, Fowler:2024}, muons are particularly problematic due to their higher energy and penetration power. As a result, they cannot be attenuated by passive shielding and represent a dominant source of irreducible correlated errors in quantum devices operated above-ground.

In this work, we explore an alternative strategy based on a cryogenic muon-tagging system that can be seamlessly integrated into existing superconducting quantum processors. The system is compatible with standard cryogenic wiring and multiplexed RF readout schemes, and can share the same data acquisition electronics typically used for qubit operation. This intrinsic compatibility allows straightforward deployment within current cryogenic setups without requiring any modification to the existing hardware.
We present the design, simulation, and experimental validation of the first prototype of such a muon-tagging system based on Kinetic Inductance Detectors (KIDs). The purpose of the present work is to demonstrate the particle-tagging capability of the system in a fully cryogenic environment, which represents a key prerequisite for its integration with sensitive quantum devices. The device operates at millikelvin temperatures and identifies muon events through coincident signals in a pair of KIDs positioned above and below the chip, respectively. We demonstrate efficient tagging of atmospheric muons ($\sim$90\%) with negligible dead time, and show that the measured coincidence rate is in excellent agreement with Monte Carlo predictions. These results pave the way for the deployment of active veto architectures in above-ground quantum computing platforms, enabling the identification and, potentially, the rejection or correction of muon-induced correlated errors in real time.

\section{System Design and Implementation}
\label{sec:system design and implementation}
The muon-tagging system is designed to identify cosmic-ray muons interacting with a superconducting quantum chip by detecting coincident signals in adjacent superconducting sensors. These signals are produced by the energy deposited in the detector substrates as the particles traverse them. The system must be compatible with cryogenic operation, minimize electromagnetic interference, and ensure both mechanical and thermal integration with the quantum device. In addition, it must cover a sufficiently large surface area to achieve high tagging efficiency while keeping the accidental coincidence rate negligible~\cite{mariani2023}.

\begin{figure}[t]
\centering
\includegraphics[width=.6\textwidth]{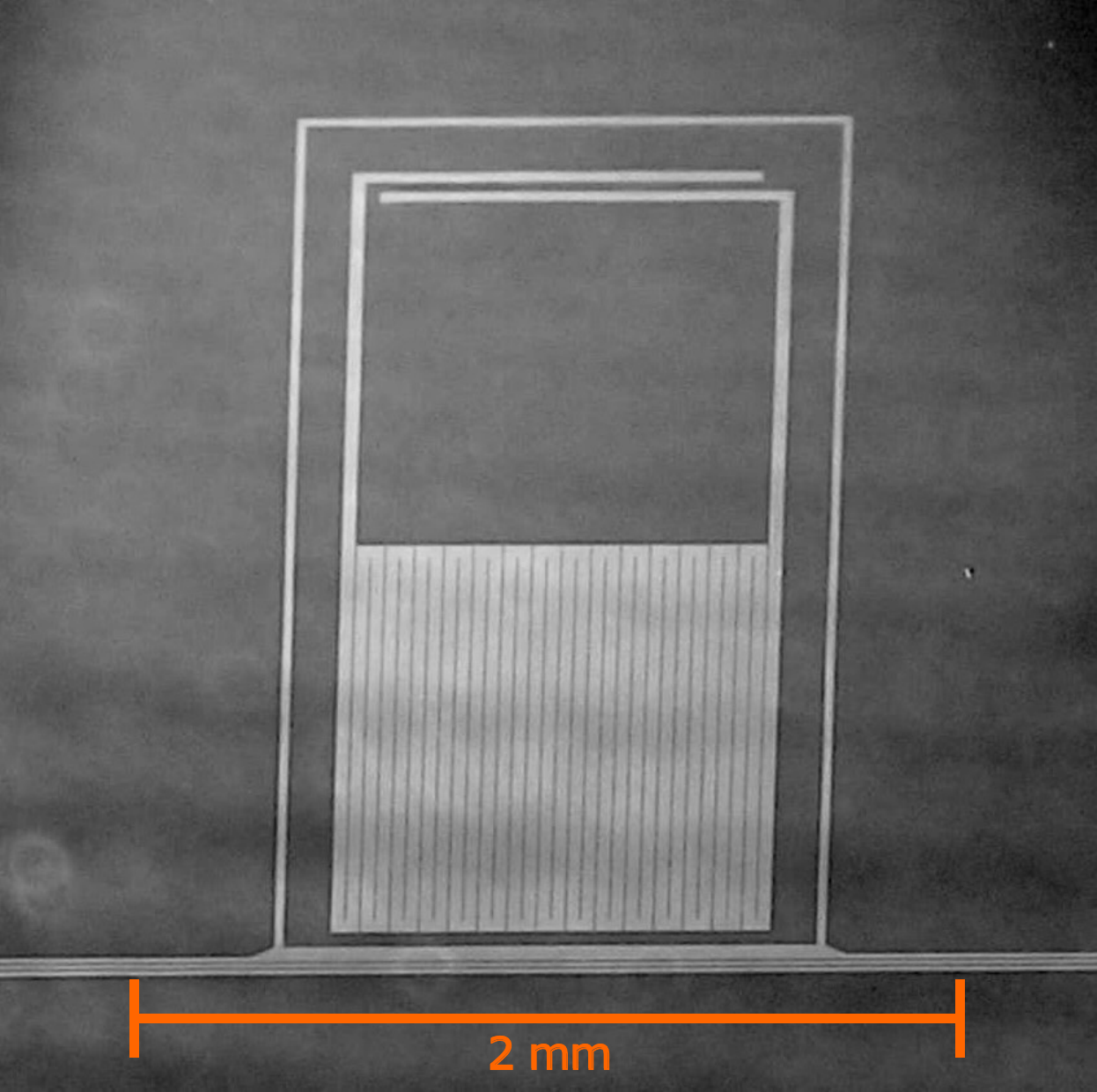}
\caption{Microscope image of a kinetic inductance detector (KID). The device consists of a meandered superconducting inductor ($\sim$6\,cm in total length and 62.5\,µm in width) connected to a two-finger interdigitated capacitor.}
\label{fig:KID_zoom}
\end{figure}

The experimental setup consists of a vertical stack of three detectors. The top and bottom layers form the muon-tagging system, while the central one acts as a proxy for the qubit chip during prototype testing. Each detector includes a 525-µm-thick, high-resistivity ($>$ 10 k$\Omega\cdot$cm) silicon substrate equipped with a KID. Each KID is a superconducting resonator consisting of a meandered inductor, approximately 6\,cm in total length and 62.5\,µm in width, connected to an interdigitated capacitor with two fingers, whose length sets the resonant frequency (Figure~\ref{fig:KID_zoom}).
All devices were fabricated by the nanofabrication team at the Superconducting Quantum Materials and Systems (SQMS) Center at Fermilab (USA). Prior to metal deposition, the native silicon oxide was removed using a buffered HF dip. A 60\,nm-thick aluminum film was then deposited on a 4-inch silicon wafer using electron-beam evaporation. The resonators, inductively coupled to a coplanar waveguide for readout, were defined by photolithography and etched into the aluminum layer using a chlorine-based dry etch. After etching, the wafer underwent standard photoresist stripping and cleaning before being diced with a precision saw according to the design layout. Each chip was then individually cleaned and packaged for integration into the experimental setup.

When a muon (or other particle) deposits energy in the silicon substrate, athermal phonons are generated and absorbed in the superconducting KID film, breaking Cooper pairs into dissipative quasiparticles. This changes the kinetic inductance, producing a shift in the resonant frequency proportional to the deposited energy~\cite{Cardani:2021_KID, Cruciani:2022}.
Thanks to the large absorber surface of the silicon wafer (20.25\,cm$^2$) compared to the active area of the KID ($\sim$3.8\,mm$^2$), the phonon-mediated detection approach provides high geometrical efficiency for muon tagging, achieving $\sim$90\% efficiency with only two devices (Sec.~\ref{sec:monte carlo simulations}). By contrast, hundreds of KIDs would be required to achieve comparable coverage around a qubit chip through a typical direct-absorption approach, where the incident particle deposits energy directly into the superconducting resonator~\cite{Adam:2018, Walter:2020}. Although the phonon-mediated approach suffers from limited phonon collection efficiency, measured to be only a few percent for aluminum KIDs with this geometry~\cite{Cardani:2015}, this limitation is not critical due to the high sensitivity of KIDs, which have energy thresholds of the order of 1\,keV, well below the average energy deposited by muons in a 525-µm-thick silicon wafer ($\approx$150 keV). Phonon-mediated detection with KIDs is therefore a promising technique and is under active development for particle detection applications~\cite{Cardani:2021_KID, Cruciani:2022, Temples:2024}.

\begin{figure}[t]
\centering
\includegraphics[width=0.48\textwidth]{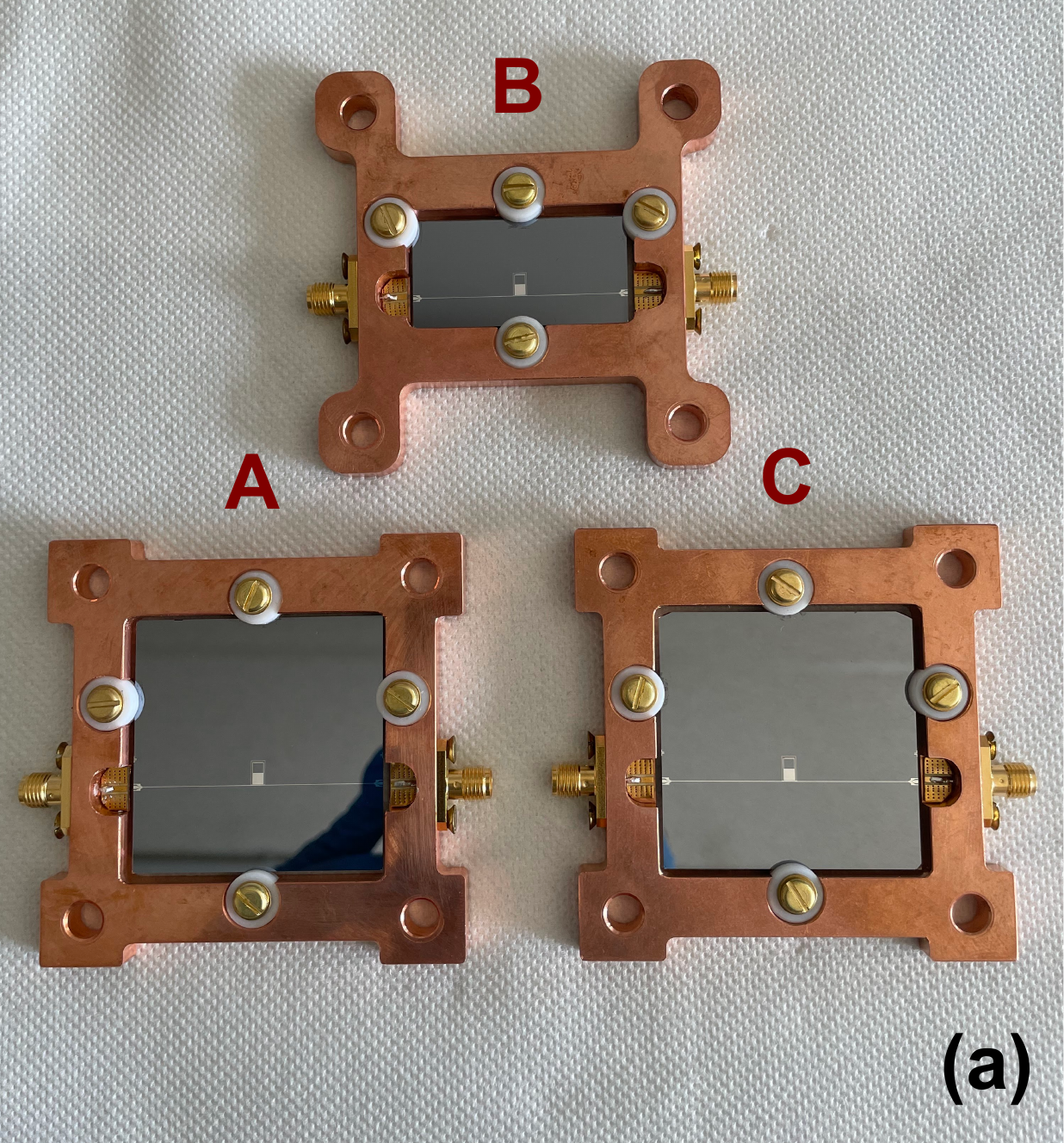}
\includegraphics[width=0.48\textwidth]{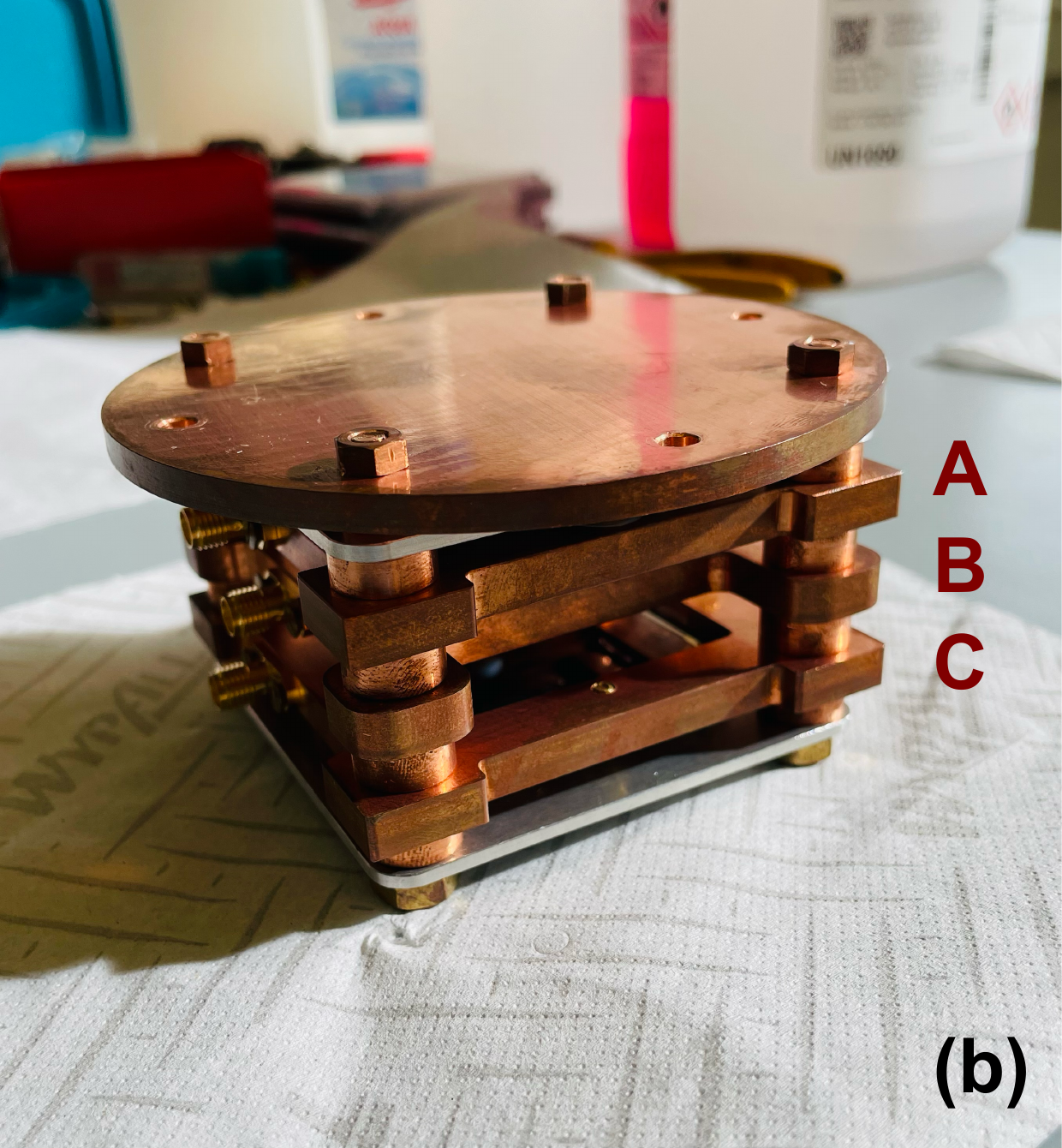}
\includegraphics[width=0.74\textwidth]{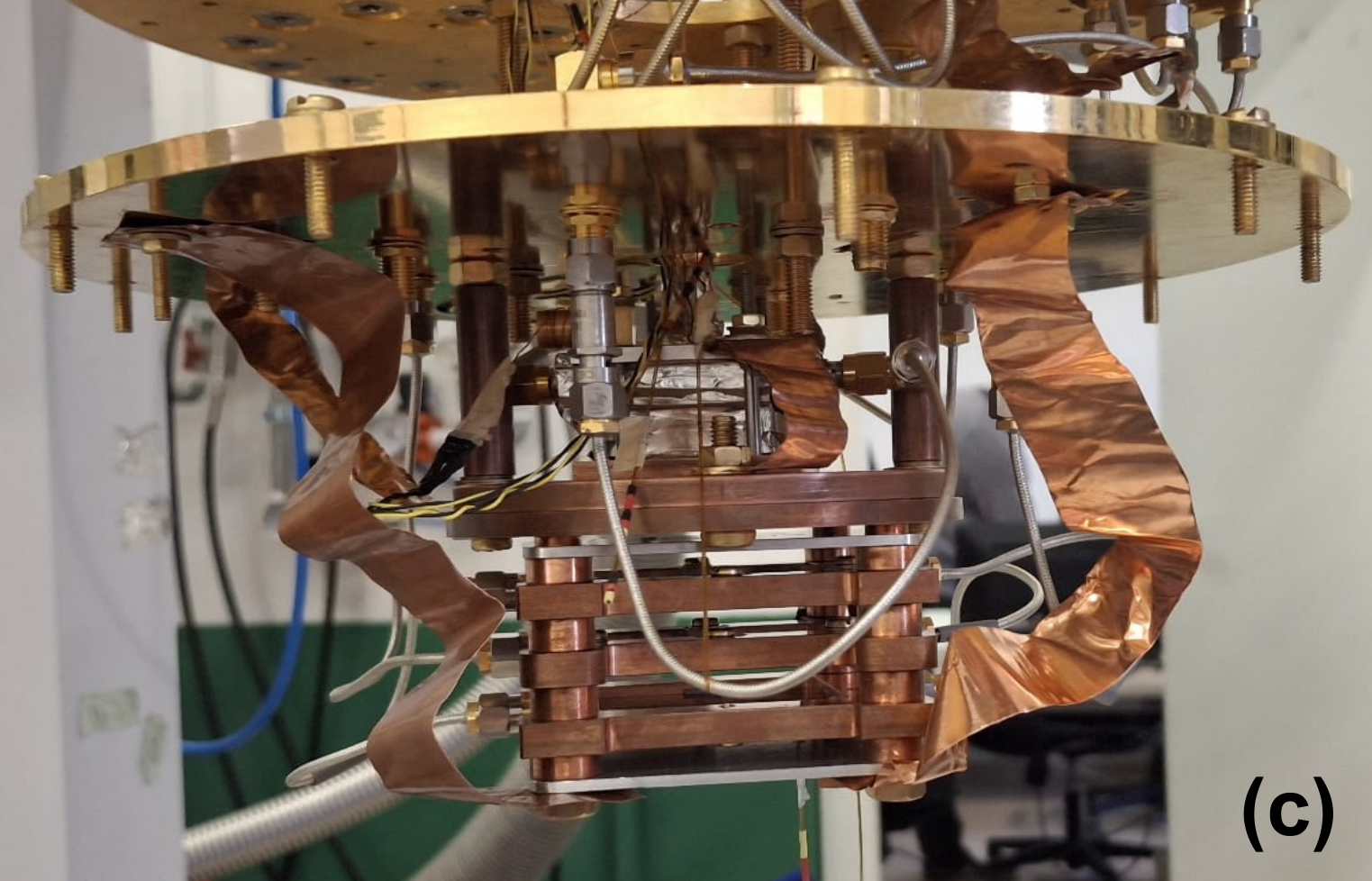}
\caption{(a) Picture of the three KID-based detectors: the top (A) and bottom (C) detectors form the muon-tagging system, while the central detector (B) serves as a proxy for the qubit chip in the prototype.
(b) Vertical stack of the detectors mounted in the copper holder, showing the top–center–bottom configuration with a 4.5\,mm inter-layer spacing.
(c) The assembled detector stack installed at the mixing-chamber stage of the dilution refrigerator at La Sapienza University.}
\label{fig:KIDs}
\end{figure}

Finally, the detectors are mounted in copper holders using Teflon supports that clamp the wafer along its edges. This configuration minimizes both mechanical stress and contact with the active area, which could otherwise introduce additional phonon-loss pathways. 
The three devices are arranged in a top–center–bottom configuration with approximately 4.5\,mm spacing between each layer. This spacing provides an effective compromise between mechanical constraints and detector proximity: it is small compared to the chip dimensions, ensuring high muon-tagging efficiency while still allowing sufficient clearance for mounting and wiring.
The top and bottom detectors, forming the muon-tagging system, have a square geometry of 4.5$\times$4.5~cm$^2$, while the central layer hosts a KID on a 4$\times2\ \mathrm{cm}^2$ substrate.
The central detector was intentionally designed with this geometry to emulate the footprint of a superconducting quantum processor. This area ($\sim$8 cm$^2$) is representative of a chip hosting on the order of 80–100 qubits, although the exact dimensions depend on the qubit layout and specific processor geometry. This choice allows testing the muon-tagging geometry under conditions representative of a real quantum device. In our applications, which employ smaller chips (typically $<$ 1\,cm$^2$), an even higher tagging efficiency is expected, since the external detectors would cover a larger fraction of the active area. Figure~\ref{fig:KIDs} shows the three detectors and their assembly in the vertical stack.

\section{Monte Carlo Simulations}
\label{sec:monte carlo simulations}
To evaluate the expected performance of the muon-tagging system and to guide the prototype design, a Monte Carlo simulation was developed using the Geant4 toolkit~\cite{AGOSTINELLI2003250}. The simulation implements the full three-dimensional geometry of the detector stack, the copper holders, and a simplified model of the cryostat including a three-layer shielding pot (Al/Cu/Cryophy\textregistered) described in Sec.~\ref{sec:experimental setup and data acquisition}. 
The three devices are modelled as 525-\,µm-thick silicon substrates with the same dimensions as the fabricated detectors, arranged in the top–center–bottom configuration illustrated in Figure~\ref{fig:KIDs}c.
In the model, the central layer, acting as a proxy for the qubit chip, is enclosed in a 5 mm-thick copper box, while the top and bottom detectors are mounted in open copper holders, reproducing the experimental setup. This configuration replicates the mechanical mounting commonly adopted for superconducting quantum processors, where the chip is enclosed in a copper box for mechanical support and thermalization. In real qubit setups, the copper box is also surrounded by a thin mu-metal shield. Since such shielding affects only the magnetic environment and not the passage of minimum-ionizing particles, it is not included in the simulation and has no impact on the expected tagging performance.

The simulation tracks the passage of particles and their energy deposition in the silicon substrates. The subsequent phonon propagation and detector response are not modelled; instead, the deposited energy is used as a proxy for the production of detectable signals. Because KIDs are sensitive to any energy deposition in the substrate, signals produced by environmental $\gamma$-rays, electrons, or $\alpha$ particles are indistinguishable from those induced by muons. At high single-detector rates, this can lead to accidental coincidences within the coincidence window, increasing the system dead time and thus reducing the effective muon-tagging efficiency.
Among these backgrounds, $\gamma$-rays dominate due to their higher flux compared to other particles~\cite{cardani:2023}, and were therefore simulated explicitly.

Primary particles, i.e., the ionizing particles injected into the simulation, were generated within the laboratory environment. Atmospheric muons were simulated to evaluate the tagging efficiency, while environmental $\gamma$-rays were included to estimate the $\gamma$-induced dead time. The muon spectrum and angular distribution reproduce measurements of cosmic-ray muons at sea level~\cite{Shukla:2018}, whereas the $\gamma$ flux was derived from measurements performed with a 3-inch portable NaI(Tl) scintillator in the laboratory where the prototype was tested (“Laboratorio di Rivelatori Criogenici” at La Sapienza University, Italy). Muons are generated from random positions on a hemispherical surface above the cryostat according to the method described in Ref.~\cite{cardani:2023}, while photons from ambient radioactivity are generated isotropically following the measured $\gamma$-ray energy distribution. Particles are then propagated through the full geometry, and the energy deposited in each silicon wafer is recorded. Interaction rates in the detectors are obtained by scaling the number of recorded events to the measured fluxes: 1 $\mu$/cm$^2$/s at sea level and 13.5 $\gamma$/cm$^2$/s in the laboratory, the latter being substantially higher than the typical 2-3 $\gamma$/cm$^2$/s observed in standard above-ground environments. 

Table~\ref{tab:MC_rates} summarizes the simulated interaction rates for muons and $\gamma$-rays in the top and bottom detectors, along with the expected coincidence rates.
For atmospheric muons, requiring simultaneous energy depositions in the two detectors yields a coincidence rate of $(195 \pm 12)\times 10^{-3}$ events/s.
For environmental $\gamma$-rays, the coincidence rate is much lower than for muons, as their interaction probability in silicon is small and the chance of producing energy depositions in both detectors is significantly reduced.
Figure~\ref{fig:MC_muons} shows the simulated energy deposition of atmospheric muons in the central chip and the subset of events tagged by the external detectors (top and bottom) forming the muon-tagging system. The ratio of tagged to total events (4589/5073) corresponds to a muon-tagging efficiency of about 90\%.

\begin{table}[t]
\centering
\footnotesize
\caption{Simulated interaction rates for muons and $\gamma$-rays in the top (T) and bottom (B) detectors, and expected coincidence rates (T–B). Rates are normalized to fluxes of 1 $\mu$/cm$^2$/s and 13.5 $\gamma$/cm$^2$/s, respectively. Uncertainties are dominated by a 5\% systematic component from the Geant4 simulation.}
\label{tab:MC_rates}
\begin{tabular}{lccc}
\hline
\textbf{Channel} & \textbf{Muon Rate} & \textbf{Gamma Rate} & \textbf{Total Rate} \\
    & \textbf{[events/s]}  & \textbf{[events/s]} & \textbf{[events/s]} \\
\hline
T     & (312 $\pm$ 19)$\times 10^{-3}$  & 2.9 $\pm$ 0.2  & 3.2 $\pm$ 0.2 \\
B  & (296 $\pm$ 17)$\times 10^{-3}$  & 3.0 $\pm$ 0.2  & 3.3 $\pm$ 0.2 \\
\hline
T–B & (195 $\pm$ 12)$\times 10^{-3}$ & (15 $\pm$ 4)$\times 10^{-3}$ & (210 $\pm$ 13)$\times 10^{-3}$ \\
\hline
\end{tabular}
\end{table}

In a coupled qubit–muon-tagging system architecture, every coincident event would activate a veto window during which quantum operations are rejected to avoid correlated errors due to muon interactions within the chip. The characteristic recovery time observed in superconducting qubits ranges from a few milliseconds~\cite{DeDominicis:2025,Harrington:2024} up to 25\,ms in some experiments~\cite{McEwen:2021}, depending on the qubit geometry. It is therefore crucial to assess whether random coincidences between independent detectors could introduce a significant fractional dead time.

Using the rates in Table~\ref{tab:MC_rates} and noting that the contribution from $\mu$–$\mu$ accidental coincidences is negligible, the accidental coincidence rate between the top and bottom detectors within a 340\,µs coincidence window (corresponding to the selection applied in the data analysis) can be estimated as the sum of:

\begin{enumerate}
    \item  $R_{\text{acc}}^{\gamma\gamma} = R_\text{T}^\gamma R_\text{B}^\gamma \Delta t \approx (3.0 \pm 0.2)\times10^{-3}$ events/s; 
    \item $R_{\text{acc}}^{\gamma\mu} = (R_\text{T}^\gamma R_\text{B}^\mu + R_\text{T}^\mu R_\text{B}^\gamma)\Delta t \approx (0.6 \pm 0.1)\times10^{-3}$ events/s; 
\end{enumerate}
The total accidental-coincidence rate is therefore
\begin{equation*}
   R_{\text{acc}} \approx (3.6 \pm 0.2)\times10^{-3} \ \ \text{events/s}. 
\end{equation*}

The simulation also predicts a true $\gamma$–$\gamma$ coincidence rate of $(15 \pm 4)\times10^{-3}$\,events/s (see Table~\ref{tab:MC_rates}).
Assuming a veto gate of 1–5\,ms, the resulting fractional dead time due to $\gamma$-induced coincidences is 
\begin{equation*}
f_{\text{DT}}^{(\gamma)} = (R_{TB}^{\gamma\gamma} + R_{\text{acc}})\tau_{\text{veto}} \approx (2\text{–}9)\times10^{-5},
\end{equation*}
i.e., well below 0.01\%. Even in the extreme case of a 25\,ms veto gate, the expected fractional dead time is 0.05\%. 
Therefore, the contribution of $\gamma$-induced coincidences to dead time is entirely negligible, confirming that ambient $\gamma$-rays do not limit the live time performance of the system, even under high-flux conditions (13.5 $\gamma$/cm$^2$/s). This makes the present estimate a conservative upper bound for the expected dead time.

These results confirm that, despite the relatively large substrate area and the resulting higher single-detector event rate, the coincidence-based muon-tagging approach is not limited by accidental or true coincidences induced by environmental $\gamma$-rays.
The simulation therefore validates the feasibility of operating such a system without significant live time losses, providing the baseline for the experimental results presented in the next sections.

\begin{figure}[t]
\centering
\includegraphics[width=1.\textwidth]{Fig3_MC_muons.pdf}
\caption{Simulated energy deposition of atmospheric muons in the central chip (blue) and events identified by the muon-tagging system (orange). The ratio of tagged to total events (4589/5073) corresponds to a muon-tagging efficiency of about 90\%.}
\label{fig:MC_muons}
\end{figure}

\section{Experimental Setup and Data Acquisition}
\label{sec:experimental setup and data acquisition}
The prototype was operated in a dry $^3$He/$^4$He dilution refrigerator at a base temperature of approximately 20 mK at Sapienza University of Rome (Italy). The vertical stack of the three detectors was enclosed in a three-layer shielding pot made of aluminum (inner), copper (middle), and Cryophy\textregistered ~(outer), which provided thermal shielding and protection from external magnetic fields. The assembly was thermally anchored to the mixing-chamber stage through copper bars for thermalization (Figure~\ref{fig:KIDs}c).

RF connectors and coaxial lines allowed the simultaneous readout of all three detectors in a daisy-chain configuration. An RF tone generated at room temperature by an Ettus X310 board was used to bias the KIDs, and attenuated by 55 dB along the cryostat to suppress thermal noise before reaching the array.
The transmitted signal was amplified by a HEMT low-noise amplifier at the 4 K stage and then returned to the same board for down-conversion.
In the context of integration with superconducting qubit platforms, the KID-based muon tagging system would operate in parallel with the qubit control and readout electronics. While the KIDs are read out here using an Ettus X310 board, the qubits could be controlled and measured using an RFSoC-based platform such as a ZCU216 board~\cite{DeDominicis:2025}. The two readout chains would initially remain independent within the same cryogenic environment. Preliminary tests with a qubit chip featuring $T_1 \sim 100\,\mu$s indicate that potential electromagnetic crosstalk between the muon-tagging system and the qubits does not introduce observable degradation of $T_1$. Synchronization between the KID and qubit acquisition systems can be achieved via optical calibration pulses injected through a fiber-coupled LED down to the mixing chamber, enabling temporal alignment of the datasets. As a first step, the veto could be applied offline by discarding qubit data within a defined time window following a KID trigger. Radiation-induced events in similar devices typically last 1--2 ms~\cite{DeDominicis:2025}, setting the relevant timescale for the veto window.

Data from the KID array were acquired using the open-source firmware developed for the BULLKID experiment~\cite{Minutolo:2019}, with a sampling frequency of 100\,kHz and a total live time of about one hour per run.
Waveforms of  fixed length (24 ms) were recorded simultaneously from all channels.
By fitting all the resonators using the method described in Ref.~\cite{Casali:2016}, we found resonance frequencies $f_0$ consistent with the design, in the 2.52--2.57 GHz range, with coupling quality factors Q$_{\mathrm{c}}\sim10^5$ and internal quality factors Q$_{\mathrm{i}}\sim10^5-10^6$, depending on the device. When an amount of energy $E$ is released into the silicon wafer, changes in the Cooper-pair density are detected by monitoring the time variation of S$_{21}$ with the RF bias set at $f = f_0$. Although both magnitude and phase of $S_{21}$ can be used, the phase typically yields a better signal-to-noise ratio.
Pulses were reconstructed with a matched filter optimized for each channel~\cite{Gatti:1986}, and only those exceeding five times the baseline RMS were retained to suppress spurious triggers. For our aluminum-on-silicon KIDs, this threshold corresponds to energies well below 1 keV, ensuring that nearly all muons deposit a detectable signal (see Fig.~\ref{fig:MC_muons}). While more refined strategies for setting the energy threshold have been proposed~\cite{Mancuso:2019}, the conventional 5×RMS criterion is robust and does not affect the observed muon rates.

\begin{figure}[t]
\centering
\includegraphics[width=1.\textwidth]{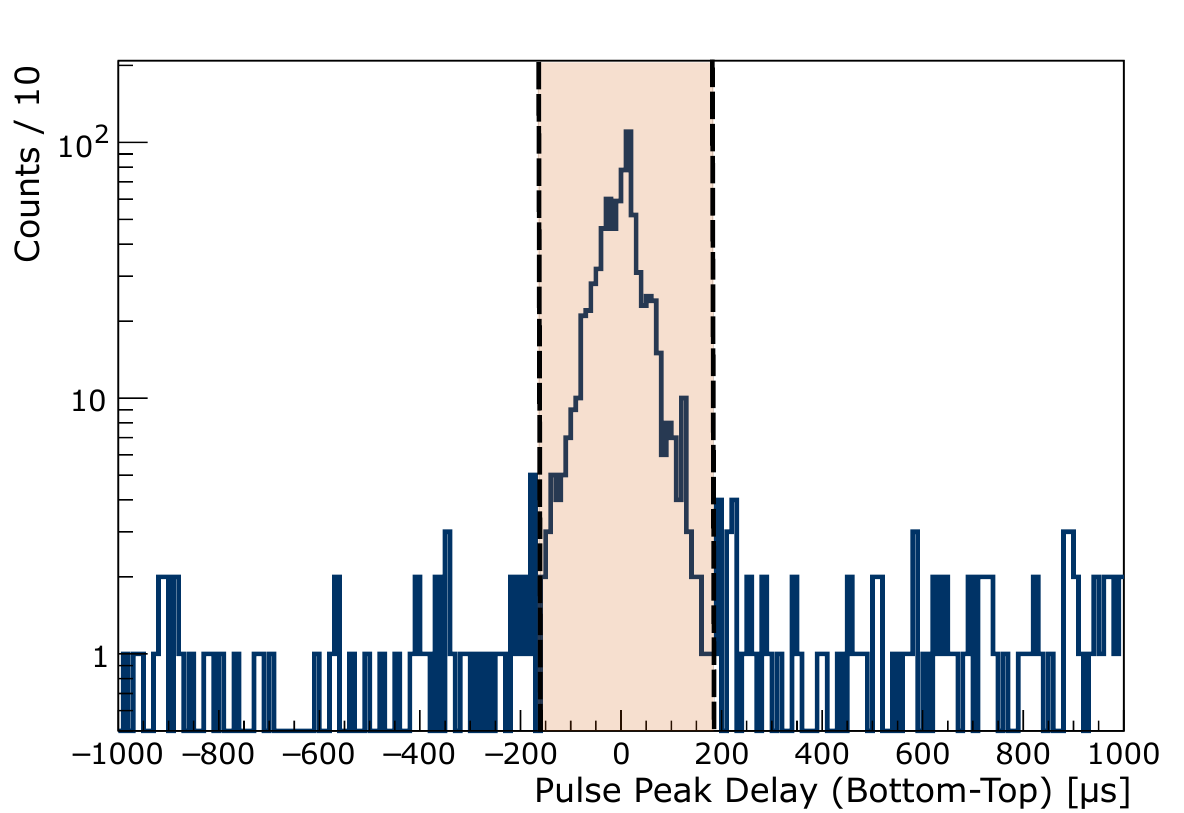}
\caption{Distribution of the \textit{Pulse Peak Delay} between the bottom and top detectors.
The shaded area ($|\Delta t| \leq 170$\,µs) represents the 340\,µs-wide coincidence window used to select true coincident events, while the flat regions at larger delays correspond to random coincidences (sidebands) employed to estimate the accidental background.}
\label{fig:concidence_window}
\end{figure}

\begin{figure}[t]
\centering
\includegraphics[width=1.\textwidth]{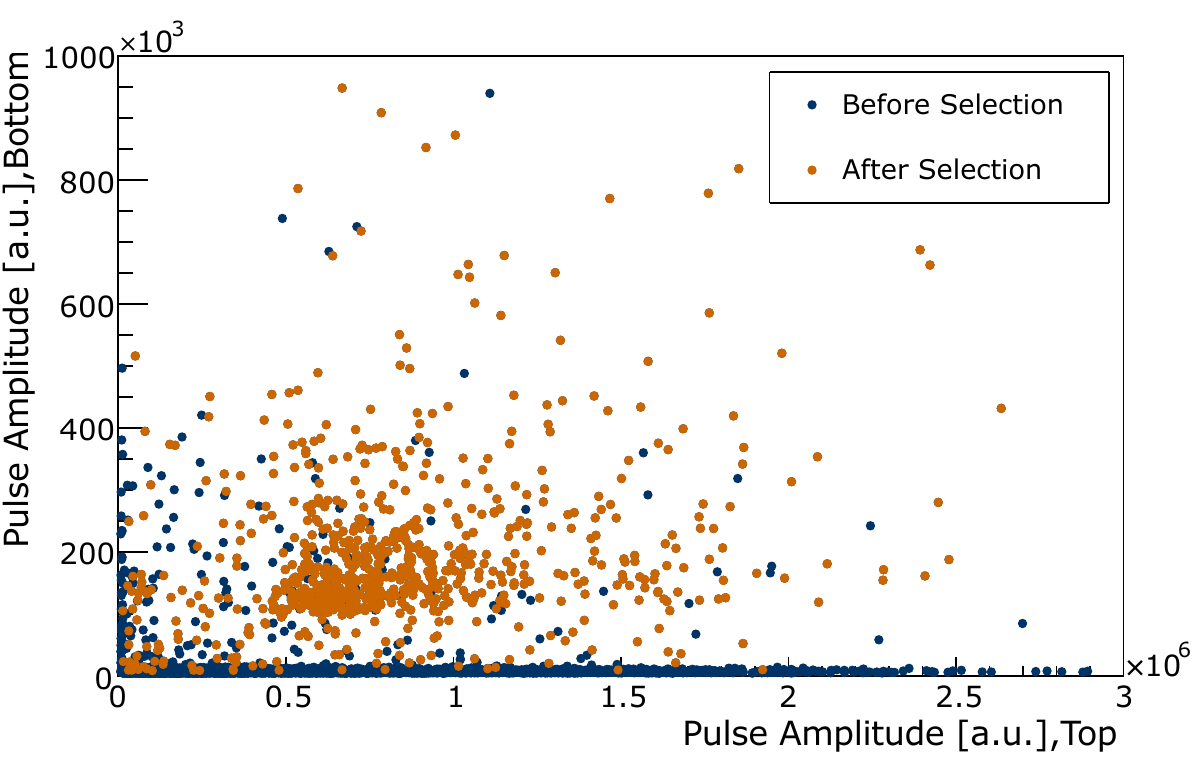}
\caption{Scatter plot of pulse amplitudes in the top and bottom detectors before (blue) and after (orange) applying the selection cuts described in the text. Before the cuts, two populations are visible: a cluster of coincident events and single-detector or uncorrelated events, which appear as negligible amplitudes in one of the detectors. After the cuts, the cluster of true coincidences stands out more clearly.}
\label{fig:scatter}
\end{figure}

\begin{figure}[t]
\centering
\includegraphics[width=1.\textwidth]{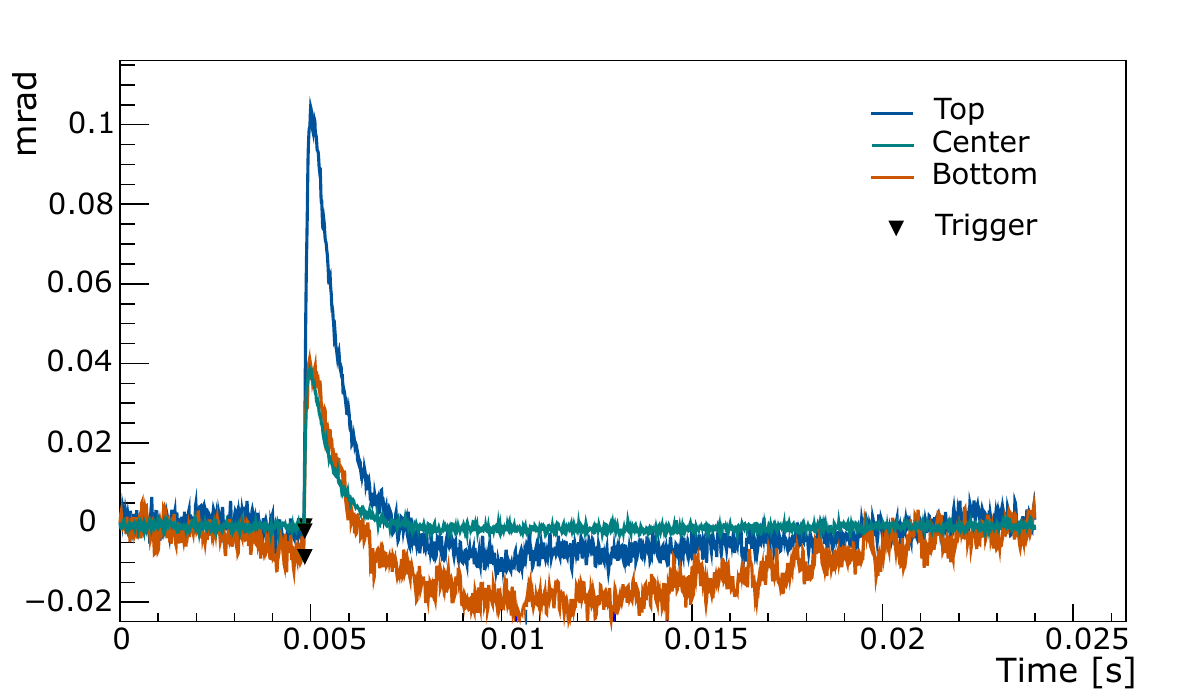}
\caption{Example of a muon event producing coincident signals in the three detectors.}
\label{fig:coincidence}
\end{figure}

Particle-induced pulses exhibited decay times of several hundred microseconds, enabling effective separation of single events and sub-millisecond time resolution for coincidence analysis.
Coincidences were identified by selecting a pulse maximum in one detector (reference channel) and verifying that the maximum in the other detector occurred within a $\pm$170\,µs time window, resulting in a total coincidence window of 340\,µs.
The distribution of time differences between the pulse maxima, defined as Pulse Peak Delay (Bottom–Top), is shown in Figure~\ref{fig:concidence_window}, where the central shaded region marks the chosen coincidence window. The $\pm$170\,µs limit corresponds to three standard deviations of the Gaussian fit to the coincidence-time peak in the data and is also consistent with the characteristic rise time of the KID pulses. Due to the finite rise time of the detectors, the temporal ordering of energy depositions in the top and bottom devices cannot be reliably inferred from the pulse maxima. The coincidence window is therefore defined to identify events occurring in both detectors, without attempting to determine their sequence.
Figure~\ref{fig:scatter} shows the pulse amplitudes in the two detectors before and after applying the amplitude and time-selection cuts. In the pre-selection sample, two populations can be identified: a cluster of coincident events, where both detectors register high-amplitude signals, and single-detector triggers, which appear as events with negligible amplitude in one of the two detectors. The latter corresponds to events interacting with a single detector, such as $\gamma$-rays. After applying the cuts, the cluster of true coincidences becomes clearly visible. 
Figure~\ref{fig:coincidence} displays an example of a muon event producing simultaneous signals in all three detectors. The three detectors exhibit different pulse amplitudes and baseline noise levels, as expected for independent KID devices due to small variations in resonator parameters, coupling conditions, and readout chain performance. While individual pulse amplitudes fluctuate on an event-by-event basis, baseline RMS values shows minor variations throughout the data-taking period, ensuring reliable operation without frequent recalibration. Event selection thresholds are defined relative to each detector’s RMS, providing a consistent signal-to-noise requirement and avoiding bias in coincidence identification. Overall, the relative performance of the detectors is similar across the dataset, supporting the robustness of the coincidence selection.

\section{Results}
\label{sec:results}

\subsection{Single-detector rates}
Applying the baseline RMS selection cut described in Sec.~\ref{sec:experimental setup and data acquisition} to the reconstructed pulse amplitudes yields the single-detector rates reported in Table~\ref{tab:measured_rates}.
The measured values are in good agreement, within uncertainties, with the simulated interaction rates reported in Table~\ref{tab:MC_rates}.

\begin{table}[t]
\centering
\footnotesize
\caption{Measured single-detector and coincidence rates after applying the baseline RMS selection cut, compared with the simulated interaction rates from Table~\ref{tab:MC_rates}. The T–B coincidence rate includes contributions from muon, $\gamma$–$\gamma$, and accidental coincidences. Measurement errors are statistical only, while simulated rates are dominated by a 5\% systematic uncertainty from the Geant4 simulation.}
\label{tab:measured_rates}
\begin{tabular}{lcc}
\hline
\textbf{Channel} & \textbf{Measured Rate} & \textbf{Expected Rate} \\
& \textbf{[events/s]} & \textbf{[events/s]} \\
\hline
T & $3.43 \pm 0.03$ & $3.2 \pm 0.2$ \\
B & $2.97 \pm 0.03$ & $3.3 \pm 0.2$ \\
\hline
T–B & $(211 \pm 8)\times 10^{-3}$ & $(214 \pm 13)\times 10^{-3}$ \\
\hline
\end{tabular}
\end{table}

\subsection{Accidental coincidences}
Using the measured single-detector rates and a 340\,µs coincidence window, the expected rate of random coincidences between the top and bottom detectors is
$R_{\text{acc}}^{\text{exp}} = R_\text{T}^{\text{meas}} \, R_\text{B}^{\text{meas}} \, \Delta t \approx (3.5 \pm 0.1)\times10^{-3}\,\text{events/s}$,
in excellent agreement with the Monte Carlo prediction ($(3.6 \pm 0.2)\times10^{-3}$\,events/s).

The rate of random coincidences was also evaluated using the flat sidebands in Figure~\ref{fig:concidence_window}, corresponding to time delays outside the coincidence region.
Because the acquisition window is asymmetric—with a longer post-trigger region (see Figure~\ref{fig:coincidence})—only the negative-delay sideband was used to avoid bias.
Integrating events in the interval $-2500 \leq \Delta t \leq -400$\,µs to reduce statistical fluctuations, and scaling to the 340\,µs coincidence window, yields
$R_{\text{acc}}^{\text{SB}} = (3.9 \pm 0.4)\times10^{-3}$\,events/s,
fully consistent with both the analytical and simulated expectations.

This analysis confirms that the accidental coincidence rate is negligible compared to the total coincidence rate, ensuring that the observed excess in the coincidence window originates from true muon events.

\subsection{Muon-induced coincidences}
Applying the coincidence requirement described in Sec.~\ref{sec:experimental setup and data acquisition} yields a total top–bottom coincidence rate of
$R_{\text{T–B}}^{\text{tot}} = (211 \pm 8)\times10^{-3}$\,events/s, as reported in Table~\ref{tab:measured_rates}.
Subtracting the expected contribution from $\gamma$–induced and accidental coincidences
($R_{\text{T–B}}^{\gamma\gamma} + R_{\text{acc}} = (18.6 \pm 4.0)\times10^{-3}$\,events/s)
gives a muon-induced coincidence rate of
\[
R_{\text{T–B}}^{\mu\mu} = (192 \pm 9)\times10^{-3}~\text{events/s},
\]
in excellent agreement with the Monte Carlo prediction of
\[
R_{\text{T–B}}^{\mu\mu,\text{MC}} = (195 \pm 12)\times10^{-3}~\text{events/s}.
\]

This correspondence confirms that the system achieves the expected muon-tagging efficiency of approximately 90\%.
Given the measured single-detector rates and coincidence window, the resulting fractional dead time due to $\gamma$-induced coincidences remains below $10^{-4}$, demonstrating that the tagging performance is unaffected by background events and that the system can operate with negligible live time losses.

\section{Conclusions and Outlook}
\label{sec:conclusions and outlook}
We have presented the design, implementation, and first operation of a cryogenic muon-tagging system based on kinetic inductance detectors, developed as a proof-of-principle platform for future integration with superconducting quantum processors. The system consists of a vertical stack of two KIDs deposited on large area substrates, operated at millikelvin temperatures and readout through a multiplexed RF architecture adapted from the BULLKID experiment. A third, smaller detector placed in the middle acts as a proxy for the qubit chip during prototype testing.

Monte Carlo simulations performed with Geant4 guided the system design and established quantitative expectations for the muon-tagging efficiency as well as for both true and accidental coincidences induced by ambient $\gamma$-rays. In practical implementations for research purposes, the central module would host a superconducting qubit chip typically with dimensions of the order of 1\,cm$^2$. Under such conditions, an even higher tagging efficiency would be achieved, since the external detectors would cover a larger fraction of the active area. In our prototype, instead, the intermediate KID was intentionally fabricated with a significantly larger area (8\,cm$^2$), comparable to that of a multi-tens-qubit processor, to emulate a conservative upper-bound scenario. Conversely, larger quantum processors could still maintain efficiencies above 90\% by extending the muon-tagging system coverage, with only a marginal increase in dead time thanks to the low background rate demonstrated here.

The fabricated prototype achieved internal quality factors up to $10^6$ and pulse decay times of several hundred microseconds, enabling sub-millisecond time resolution for coincidence analysis.
Using the selection criteria described in Sec.~\ref{sec:experimental setup and data acquisition}, we measured a top–bottom muon coincidence rate of $(192 \pm 9)\times10^{-3}$\,events/s, in excellent agreement with the Monte Carlo prediction. The corresponding muon-tagging efficiency is about 90\%, with a fractional dead time below $10^{-4}$, confirming that ambient $\gamma$-rays do not limit the live time performance of the system.
 
These results demonstrate the feasibility of operating a high-efficiency, low-dead-time muon-tagging system at millikelvin temperatures, establishing the key technological prerequisite for its integration into above-ground superconducting quantum processors.
Future developments will focus on coupling the tagging system to a multi-qubit chip, introducing veto strategies or correction algorithms against muon-induced correlated errors, and optimizing KID geometries and materials (e.g., inductor volume, capacitor design, feedline coupling) to further improve detection efficiency.

\section*{Data Availability}
Data will be made available upon reasonable request.\\

\section*{Acknowledgements}
The authors thank M.~Iannone and M.~Zullo for the technical drawings and construction of the device, A.~Girardi for the technical support, and E.~Vázquez-Jáuregui for deriving the environmental $\gamma$-ray flux from the measurements performed in the “Laboratorio di Rivelatori Criogenici” at Sapienza University.

The work was supported by the INFN National Scientific Committee 5 grant for young researcher call No. 25559/2023 assigned to the ACE-SuperQ (Abatement of Correlated Errors in Superconducting Qubits) project, by the Italian Ministry of Foreign Affairs and International Cooperation, grant number US23GR09, by the U.S. Department of Energy, Office of Science, National Quantum Information Science Research Centers, Superconducting Quantum Materials and Systems Center (SQMS), under Contract No. 89243024CSC000002, by the PNRR MUR project CN00000013-ICSC, and by the Italian Ministry of Research under the PRIN Contract No. 2022BP4H73 (Chasing Light Dark Matter with Quantum Technologies). \\


\section*{Competing Interests}
The Authors declare no Competing Financial or Non-Financial Interests.

\onecolumngrid
\section*{References}

\twocolumngrid
\bibliography{arxiv_v4}

\end{document}